\documentclass[12pt,preprint]{aastex}
\usepackage{amssymb}
\def \etal{{et~al.\null}}
\newcommand{\lea}{{\la}}
\newcommand{\gea}{{\ga}}

\shorttitle{Distributions of Star Clusters in M83}
\author{Rupali Chandar,\altaffilmark{1} 
Bradley C.\ Whitmore,\altaffilmark{2} 
Daniela Calzetti,\altaffilmark{3} and
Robert O'Connell\/\altaffilmark{4}
}

\shortauthors{Chandar et al.}
\email{Rupali.Chandar@utoledo.edu}
\altaffiltext{1}{Department of Physics \& Astronomy, The University of Toledo, Toledo, OH 43606}
\altaffiltext{2}{Space Telescope Science Institute, Baltimore, MD, USA}
\altaffiltext{3}{Dept.\ of Astronomy, University of Massachusetts, Amherst, MA 01003, USA}
\altaffiltext{4}{Dept. of Astronomy, University of Virginia,  Charlottesville, VA 22904-4325, USA}

\begin{document}

\title{Star Cluster Mass and Age Distributions of Two Fields in M83 Based on 
\emph{HST}/WFC3 Observations}

\begin{abstract}
We study star clusters in two fields in the nearby spiral galaxy M83 
using broad and narrow band optical imaging taken with the WFC3 
on-board \emph{HST}. We present results based on several
different catalogs of star clusters in an inner and outer field,
and conclude that different methods of selection
do not strongly impact the results,
particularly for clusters older than $\approx 10$~Myr.
The age distributions can be described by a power law, $dN/d\tau \propto \tau^{\gamma}$,
with $\gamma \approx -0.84 \pm 0.12$ in the inner field,
and $\gamma \approx -0.48\pm0.12$ in the outer field for $\tau \gea 10$~Myr.  
We bracket the difference, $\Delta \gamma$, between the two fields 
to be in the range 0.18--0.36, based on estimates of the relative
star formation histories.
The mass functions can also be described by a power law,
 $dN/dM \propto M^{\beta}$,
with $\beta\approx -1.98 \pm0.14$ and $\beta\approx -2.34\pm0.26$
in the inner and outer fields, respectively. We conclude that the shapes 
of the mass and age distributions of the clusters in the two fields
are similar, as predicted by the ``quasi-universal'' model. 
Any differences between the two fields 
are at the $\approx$2--3$\sigma (\approx$1--2$\sigma)$
level for the age (mass) distributions. Therefore any dependence of 
these distributions on the local environment is probably weak.
We compare the shapes of the distributions with those predicted by 
two popular cluster disruption models, and find that both show evidence
that the clusters are disrupted at a rate that is approximately independent of
their mass, but that the observational results do not support  the earlier
disruption of lower mass clusters relative to their higher mass counterparts.
\end{abstract}

\keywords{galaxies: individual (M83) --- galaxies: star clusters --- stars: formation}

\section{Introduction}
The age and mass distributions of a population of star clusters 
provide important clues to their formation and disruption.
Recent studies of the cluster populations in nearby star-forming
galaxies have suggested that the observed distributions are dominated
by the disruption, rather than the formation, of the clusters.
Some physical processes, for example the evaporation of stars due to two-body
relaxation, are known to disrupt low-mass clusters earlier
than high-mass clusters, thereby imprinting features (such as a bend) on
the mass and age distributions. 
Other processes are believed to operate (approximately) independently of
the mass of the clusters, preserving the shape  
of the mass function over time and that of the age distribution
at different masses (e.g., Fall \& Zhang 2001).

There are still relatively few galaxies with well determined
cluster mass and age distributions.
Furthermore, the results from different groups working on the
same galaxy have not yet converged.
Some studies (e.g., Whitmore et~al.\ 2007, Fall et~al.\ 2009, Fall \& Chandar 2012)
advocate a ``'quasi-universal'' model for cluster populations in star-forming
galaxies, where the shapes of the cluster mass and age distributions are 
independent of one another, 
because clusters are disrupted at a rate that is approximately independent of their mass.
Other studies (e.g., Boutloukos \& Lamers 2003, Bastian et~al.\ 2012) suggest that
there is a characteristic timescale for cluster disruption processes 
that depends on the mass of the clusters, 
and this timescale depends strongly on the local environment.

The motivation for this work is three-fold: to empirically
determine the age and mass distributions of star
clusters in the nearby spiral galaxy M83, to understand
how the selection of the clusters affects the results,
and to test whether or not clusters are disrupted at a rate that depends
on their mass.

We use multi-band observations of two fields in M83,
taken as part of our program GO-11360, for this purpose. 
These images were observed with the WFC3 camera
on-board the \emph{Hubble Space Telescope} (\emph{HST}).  
We present new cluster catalogs selected using
different methods and compare with catalogs previously
published by Chandar et~al.\ (2010) and Bastian et~al.\ (2012).
The cluster age and mass distributions observed
in each field are then compared with predictions from
mass-dependent and mass-independent disruption models.

The rest of this paper is organized as follows.  
 Section~2 summarizes the observations and photometry,
describes the selection procedures used to
create the various star cluster catalogs, and 
explores the similarities and differences between these
catalogs. This section also summarizes the method used to
determine the ages and masses of the clusters.
Section~3 presents the age and mass distributions of the
clusters in the two fields in M83 using all of the cluster catalogs.
Section~4 discusses the implications of the mass-age results from the two fields
for the formation and disruption of the clusters.
We summarize the main results of this work in  Section~5.

\section{Observations, Photometry, and Cluster Catalogs}

\subsection{\emph{HST} Data \& Photometry}

Multi-band images of two $3.6\times3.6$~kpc$^{2}$ portions of M83
were taken with the WFC3 on \emph{HST},
one covering the nucleus (inner field) and presented previously in
Chandar et~al.\ (2010), and a second pointing to the north (outer field).
These observations were taken as part of the Early Release Science program~1  
(GO-11360; PI: O'Connell).
In this work, we use observations taken in five filters,
$F336W$ ($U$), $F438W$ ($B$), $F555W$ (inner field) or $F547M$ 
(outer field) ($V$), $F814W$ ($I$), and $F657N$ (H$\alpha$).
Three or four separate exposures were taken at slightly offset positions
for each filter, in order to improve the spatial resolution.
Figure~1 shows color images of each field.

The flat-fielded WFC3/UVIS images in each filter were co-added using
the MULTIDRIZZLE task (Koekemoer et~al.\ 2002), with a final 
pixel scale of $0.0396 \arcsec$~pixel$^{-1}$.
These reduced images are available at:
{\tt http://archive.stsci.edu/prepds/wfc3ers/m83datalist.html}.
As before, we assume a distance of $4.5\pm0.2$~Mpc to M83 (Thim et~al.\ 2003),
corresponding to a distance modulus of $(m-M)_0 = 28.28 \pm 0.1$, and a 
pixel scale of $0.876$~pc~pixel$^{-1}$.

Detection and photometry of the sources in the data
were performed as described in Chandar et~al.\ (2010).
In order to detect as many sources as possible,
we aligned and co-added together 
the final, drizzled broad-band images in
the $F336W$, $F438W$, $F555W$, and $F814W$ filters, 
normalized by the typical rms in each image.
This results in a ``white-light" image which gives
approximately equal weight to the different filters.
This procedure allows us to include objects
in our source list that are very blue or very red,
such as blue and red supergiant stars, that 
otherwise might not be detectable in any given filter.
We identified all sources, both point-like and slightly extended,
using the IRAF task DAOFIND on a median-divided
white light image (see discussion in Miller \etal\/ 1997).
This initial source catalog contains 
individual stars, close blends of a pair or a few stars,
star clusters, and background galaxies.

We perform circular aperture photometry of all
detected sources on the drizzled
images for each filter using the IRAF task PHOT
with an aperture radius of 3~pixels
and a background annulus between 10 and 13~pixels.
For the narrow-band F657N (H$\alpha$) image, 
we perform photometry on the original image without subtracting
the stellar continuum flux. Aperture corrections were made based on the
measured size of each cluster, using the formula
given in Chandar et~al.\ (2010).
We convert the instrumental magnitudes to
the VEGAMAG magnitude system by applying the following
zeropoints: $F336W = 23.46$, $F438W =  24.98$, $F555W =  25.81$,
$F657N = 22.35$, and $F814W = 24.67$, which are
provided by STScI at the following URL:
{\tt http://www.stsci.edu/hst/wfc3/phot/zp/lbn}.
We will loosely refer to these as ``$U$,''``$B$,''
``$V$,'' ``H$\alpha$,'' and ``$I$''-band magnitudes, although we do not
make any transformations to the Johnson-Cousins system.

\subsection{Cluster Selection and Comparison Between Catalogs}

One of the goals of this work is to assess the impact that the
method used to select stellar clusters has on the resulting
mass and age distributions. Bastian et~al.\ (2012) discussed the impact of 
selection by comparing their results with those presented in Chandar et~al.\ (2010)
for the inner field, but this is the first time that such a comparison
has been made for the outer field in M83.

There are three methods of cluster selection that are discussed below:
(1)~fully automated selection based on criteria that 
find objects that are slightly broader than the PSF, followed by minimization
of blends using algorithms discussed in Chandar et~al.\ (2010), 
(2)~fully manual selection, based on a detailed by-eye examination of the sources
within the images to assess if they are broader than the PSF (Chandar et~al.\ 2010), 
and (3)~a ``hybrid'' method where contaminants are rejected based on
visual inspection after generating an automated catalog (Bastian et~al.\ 2012).  
While automatic methods are preferred in terms of repeatability and objectivity, these methods often have difficulty distinguishing close pairs of stars from resolved star clusters, especially in crowded regions. Although manual methods are subjective in nature, they have the advantage that each candidate is examined individually.  We find many cases 
which are easily classified by eye but incorrectly identified by automated techniques
(e.g., a diffuse cluster with a very nearby star or diffraction spikes from saturated stars).
For these reasons, we construct both manual and automatic catalogs, and use the differences between them to help estimate uncertainties in the resulting mass and age distributions. 

One of the key differences between the Bastian et~al.\ selection and
ours is that they attempt to make a distinction between bound and
unbound clusters, by eliminating sources that are asymmetric or do not
have a strong central concentration. Effectively, this eliminates
mostly very young $\tau \lea 10^7$~yr clusters, such as source 5652
in Figure~2, which is slightly asymmetric, but appears to be a bonafide, 
compact star cluster. We do not attempt to make a distinction between 
bound and unbound clusters based on morphology, 
since it is not possible to tell from the appearance of a cluster whether it has positive
or negative total (potential plus kinetic) energy
(e.g., Baumgardt \& Kroupa 2007).
Even with velocity dispersion measurements made from high resolution
spectroscopy it is difficult to determine if a cluster is bound or unbound.

In this work, we use six catalogs of compact star clusters,
three each in the inner and outer fields, selected using the
methods described above.
The total number of clusters in each catalog are:
\emph{(i)} {\bf Manual:} $\approx$490 inner field, $\approx$530 outer field;
\emph{(ii)} {\bf Automatic:} $\approx$660 inner field, $\approx$720 outer field; and
\emph{(iii)} {\bf Hybrid:} $\approx$460 inner field, $\approx$480 outer field.

The top panel of Figure~2 compares the three cluster catalogs
in the portion of the outer field outlined in Figure~1.
This figure shows several interesting results.
Our automatic catalog misses very diffuse clusters, and includes
as clusters some questionable sources in the most crowded regions.
Our manual catalog appears to do significantly better, but it is subjective
and therefore difficult to assess completeness quantitatively.
The Bastian et al. catalog has few sources in crowded regions,
and appears to go deeper in regions of low background.
The lists of cluster candidates in this area agree at about the 40--70\%
level, similar to the $\approx$60\% level of agreement
found by Bastian et~al.\ (2012) for the catalogs covering the inner field.
The reader is referred to Whitmore et~al.\ (2013) for more detailed discussion,
including a comparison with a fourth catalog obtained using automatic
criteria on Hubble Legacy Archive ({\tt http://hla.stsci.edu}) source lists,
which also agrees at roughly the 60\% level.
The key question, however, is whether or not these different cluster catalogs
based on different selection criteria yield strongly
different mass and age distributions.

\subsection{Age Dating Technique}

Figure~3 compares the $U-B$ vs.\ $V-I$ colors measured for the clusters in each catalog with
predictions from the twice
solar metallicity stellar population models of Charlot \& Bruzual (2009;
hereafter CB09, private communication; see also Bruzual \& Charlot 2003).
The model predictions, shown as the solid line, cover the age range from $10^6$ 
to $10^{10}$~yr, starting from the upper left.
The measured colors match the model predictions fairly well,
despite the fact that no correction has been made for any reddening
due to dust. Reddening of individual clusters will `smear' out the distribution 
along the indicated vector.
We see that the distributions in color-color space are similar amongst
all three catalogs, which show a higher relative concentration of clusters with blue colors
in the inner field than in the outer field.

We estimate the age $\tau$, extinction $A_V$, and the mass for each cluster as we have
done in previous works (see e.g., Fall et~al.\ 2005; Chandar et~al.\ 2010 for details), by
comparing the observed and predicted $U$, $B$, $V$, $I$, and H$\alpha$ magnitudes.
We perform a $\chi^2$ fit to the predictions from the CB09 stellar population
models with solar metallicity. The narrow-band H$\alpha$ filter contains both stellar continuum and nebular line emission, allowing us to include it directly in the fit for 
clusters of all ages. The mass of each cluster is estimated from the extinction-corrected 
$V$ band luminosity, assuming a Galactic extinction law (Fitzpatrick 1999),
and the age-dependent mass-to-light ratios ($M/L_V)$ predicted by the CB09 models.
We assume a Chabrier (2003) initial stellar mass function; if we had adopted the Salpeter (1955) rather than the Chabrier IMF, the $M/L_V$ and hence the estimated masses would increase by a near constant $\approx$40\%.

We note that the predictions from the BC09 models assume that the stellar IMF for 
each cluster is fully sampled.
The colors of clusters with masses below   $\approx$10$^4~M_{\odot}$ 
begin to spread significantly around these predicted models, because the upper end 
of the stellar IMF within these clusters are not fully populated.
These stochastic fluctuations in the number of massive stars can lead to variations in
the intrinsic optical colors of clusters with similar ages (and masses).
However, Fouesneau et~al.\ (2012) found that the
resulting age and mass distributions are similar for clusters more
massive than a few $\times10^3~M_{\odot}$ regardless of whether the observed colors
are compared with fully sampled or stochastic model predictions. Using this as a guideline, we use $\approx$$3 \times 10^3~M_{\odot}$ as a lower limit for our analysis.

\section{Results}

\subsection{Cluster Mass vs.\ Age Diagrams}

In Figure~4 we show the log~$M$-log$\tau$ diagrams for the different cluster
catalogs in the inner (upper panels) and outer (lower panels) fields observed in M83.
The method used to determine the cluster masses and ages for the catalogs presented here 
is described in Section~2.3. We do not rederive the masses and ages of clusters in the Bastian et~al.\ (2012) catalogs, but use their values directly.

The solid (approximately) diagonal line in each panel shows predictions for a $M_V=-6.0$ cluster in the mass-age plane, and is a reasonable estimate of the magnitude limit of our
cluster catalogs. A comparison of the results shows that, at least qualitatively, the three different catalogs provide similar distributions in the inner and outer fields. 
The $M-\tau$ diagrams show that clusters in our fields have formed with a continuous
range of ages from $\approx$$10^6$ to $10^{10}$~yr. The diagrams also show a number of small-scale features, with gaps at some ages and apparent over-densities of clusters at other ages.  These types of features are expected when estimating ages by comparing observed
 cluster spectral energy distributions with predicted stellar evolution models, because
 the predicted colors loop back on themselves, creating a fairly small range in the 
 predicted shapes of the spectral energy distributions over a relatively large age span, 
 resulting in gaps and other small-scale artifacts.  These small-scale features do not however, affect the broad distribution of cluster  masses and ages in a significant way (e.g., Fouesneau et~al.\ 2012).

\subsection{Cluster Age Distributions}

The cluster age distribution, $dN/d\tau$, provides important information about the formation and disruption of the clusters. We determine this distribution by simply counting clusters above a given mass limit,  in equal size bins of log~$\tau$. The age distributions determined from each cluster catalog are plotted for two different ranges of 
mass, selected to stay above the approximate magnitude limit shown
by the diagonal fading track, and also to masses higher than log~$M\approx 3.5$
(where stochastic fluctuations do not have a significant affect on the distributions;
see Section~2.3). In order to make a fair comparison between all three catalogs,
we do not plot clusters younger than 10~Myr in this particular diagram, 
\textbf{since the Bastian et~al.\ (2012) catalogs are likely incomplete
at these very young ages.} We will present the age distributions of clusters over the entire age range for all seven fields observed in M83 in a future work.
We use a typical bin width of $\approx$0.4--0.5 in log~$\tau$ in order to 
have a sufficient number of bins, at the expense of being able to fully
bin over gaps and other small-scale features resulting from the age dating procedure.

Figure~5 shows the resulting age distributions for two different mass-limited
samples determined for each catalog. Each age distribution has a steadily declining shape, with no obvious bends or other features, except possibly for $\tau \lea 10^7$~yr in a few cases.
The shapes of the cluster age distributions in each field are approximately the same
for both plotted intervals of mass. This indicates that the age distribution is independent
of the mass of the clusters, at least over the range studied here. The age distributions can be approximated by power laws of the form $dN/d\tau \propto \tau^{\gamma}$.
We perform a least squares fit to log$(dN/d\tau) = \gamma$log$~\tau +$ const, and
present the best fit values of $\gamma$ in Figure~5 and compile them in Table~1.
The age distributions determined from the different catalogs of  the inner field are all similar, 
with a mean $\gamma = -0.84$. The outer field distributions give a mean of $\gamma =  -0.48$. If we include clusters younger than 10~Myr, we find slightly steeper values for
$\gamma$ in both fields (for the inner field, see results presented in Chandar et~al.\ 2010).

The relative age distributions can be affected by differences in the star formation histories between the two fields, as also pointed out by Bastian et~al.\ (2009, 2012).
We make a first attempt to correct for this difference by using the 
relative star formation rates for the inner and outer M83 fields determined by 
Silva-Villa et~al.\ (2013) from an analysis of the colors and luminosities of individual stars
in \emph{HST}/ACS observations taken at somewhat different locations within M83,
and presented in Figure~11 of Bastian et~al.\ (2012). This figure shows that the star formation rate in the inner field is approximately a factor of three
higher than in the outer field between the ages of  log$\tau=7.0$--7.6, but that this ratio
appears to decline to a factor of 2 at an age of log$\tau=8.0$.
Making a simple relative correction of log~$(3/2) = 0.18$ to $\gamma$
would result in the $\gamma$ values in the two fields becoming more similar, with $\Delta \gamma = 0.18$. However, given the large uncertainties in  the determination of the relative star formation histories between the two fields over the studied range of ages, it is also possible that there is little difference between the two. Hence, we bracket the difference in $\gamma$ between the two fields to be within the range $\Delta \gamma = 0.18 - 0.36$.

In order to assess whether or not this is a statistically significant difference,
we need to estimate the uncertainties on the values of $\gamma$.
Previous experiments have shown that the \emph{absolute} uncertainty is
typically 0.2 (e.g., Chandar et~al.\ 2010; Fouesneau et~al.\ 2012).
However, we are more interested here in the \emph{relative} uncertainties rather
than the absolute ones. In this case, many of the systematic differences, which dominate the absolute uncertainties (e.g., use of different model predictions, age dating methods, etc), 
tend to cancel out. Other systematic differences, such as those in the relative star formation histories between the two fields, will still be present. The scatter between the six different fit values in Table~1 for each field is $\approx$0.10, and the error in the mean is therefore 
$\approx$0.04 (i.e., scatter/$\sqrt{N}$). This is probably an underestimate, however, because
the cluster samples are not fully independent but overlap by $\approx$60\%, and so dividing by $\sqrt{N}$ is not fully justified. Any difference in  star formation history between the two fields will introduce an additional uncertainty. The true relative error is likely to be somewhere between 0.04 and 0.20, and we adopt the average value of $\approx$0.12, 
which is similar to the scatter in the mean,  as the uncertainty.
\textbf{Any difference in the age distribution between the inner and outer fields is therefore tentative.} The addition of five more fields recently observed with \textit{HST}/WFC3
(proposal = 12513, PI = Blair) will provide a more definitive test in the future. 

\subsection{Cluster Mass Functions}

The shape of the cluster mass function, $dN/dM$, is one of the key
diagnostics of whether cluster disruption is dependent or independent
of mass. We determine these distributions by counting clusters in different
intervals of age, and restricting the low mass end to stay above the
magnitude limit shown in Figure~4. The specific age intervals are: 
(1) log~$\tau=6.0$--7.0 (blue),
(2) log~$\tau=7.0$--8.0 (green), and
(3) log~$\tau=8.0$--8.6 (red).
Because the small-scale features observed in the mass-age diagram
have less impact on the mass function, we plot $dN/dM$ using
equal numbers of clusters in each bin of log~$M$.
In Chandar et~al.\ (2010) we showed that the particular method of binning,
whether equal in width or in cluster number, only affects the value of 
$\beta$ at the $\pm0.05$ level, smaller than the actual uncertainties.

Figure~6 shows the mass functions resulting from the three different catalogs
(our manual, our automatic, and Bastian et~al.\ hybrid) for both the inner (upper panels) and outer (lower panels) fields, in the three different intervals of age listed above.
Each of these distributions can be well-represented by a simple power-law.
We perform fits of the form log~$dN/dM = \beta$~log~M + const, show the
best fit as the solid line in each figure, and record the value of $\beta$ in Table~2.

The inner field mass functions have a mean of $\beta =  -1.98$, and the outer field distributions give $\beta = -2.34$. Absolute uncertainties for $\beta$, based on experiments with different age dating methods, filter sets, binning, etc.\ are $\approx 0.2-0.3$ (see Chandar et~al.\ 2010). The scatter between the nine different fit values in Table~2 is 0.20 for
the inner field and 0.39 for the outer, and the uncertainty in the means are 0.07 and 0.13, respectively. For the uncertainty in $\beta$ for each field, we adopt the mean value between the scatter and the uncertainty in the mean, resulting in $-1.98\pm 0.14$ for the inner field and $-2.34\pm 0.26$ in the outer. If the most discrepant value of $\beta$ (i.e. lower-right panel in Figure~6) in the outer field is excluded, we find $\beta$ of $-2.44 \pm0.19$.

Therefore, while the mass function may be somewhat steeper in the outer field when compared with the inner field, this result is only significant at the $\approx$1--$2\sigma$ level. Just as for the inner field, we find no significant change in the shape of the cluster mass function in the outer field when going  from youngest to oldest. \emph{This is one of our key results, that the shapes of the cluster mass functions in both the inner and outer fields are similar at different ages, and show no systematic flattening, particularly at the low mass end, from youngest to oldest.} In order to better demonstrate this result, we show the mass functions normalized to lie on top of one another in Figure~7. This figure shows clearly that, although different ranges of mass are plotted in each age interval, the shapes are essentially the same within the uncertainties, and the oldest (red) clusters at the low mass end do not fall below the dashed lines.

Perhaps the strongest deviation in shape is the slightly steeper slope for the oldest clusters 
in the upper two left panels (red points in Figure~7). We note that this is opposite to the 
trend predicted by a short disruption time scale inferred for the inner field by Bastian et~al.\  (2012), as will be discussed further in \S4.3.  We also note that the intermediate interval of age, log~$\tau=7$--8, particularly in the outer field for our manual and the Bastian catalogs,
appear to be somewhat flatter than in the other age ranges. This effect may be related to
the age-dating artifact mentioned in Section~3.1.

 Bastian et~al.\ (2012) suggested that the cluster mass function in the two fields  is inconsistent with a simple power-law, rather that it requires a  Schechter-like cutoff at the high mass end, where $dN/dM \propto M^{\beta} exp{(-M/M_C)}$.  In order to assess whether or not this is the case,  we first examine the mass functions of 100--400~Myr-old clusters  from the three different catalogs.   The mass functions for clusters in the inner field, shown in Figure~6,   are quite consistent with a pure power-law in all three catalogs.  In the outer field there is a hint of a Schechter-like cutoff at the  high mass end in the Bastian et~al.\ catalog, but it is not  statistically significant ($\lea2\sigma$).  We do not see a similar feature in our best catalog, the manual catalog presented here.  Next, we experimented with cumulative mass functions in the different  age intervals and different catalogs with simulated ones drawn from  a single power law. We find similar results to the binned case, 
 that the simulated and observed cumulative mass functions  match well, if values of $\beta$ similar to those compiled in Table~2 are  used. The one exception is  the mass function for 100--400~Myr clusters in the outer field, which appears  to be slightly deficient in massive clusters.  The Bastian et~al.\ (2012) conclusion is based on a comparison of the cumulative distribution of cluster masses with those for a pure power law and Schechter functions
with different values for $M_C$. However, for this test, shown in their Figures~15 and 16, they have required an exact value of $\beta=-2.0$. They have also used a different interval of cluster ages, from 3--100 Myr, than in their subsequent analysis (e.g., in their Figure~17),
Incompleteness in their cluster catalog below ages of 10~Myr may, however, bias their result.
In any case, the number of clusters involved is only a few out of a sample of a few  hundred.

\section{Discussion}

\subsection{Predictions of Cluster Disruption Models}

There are currently two popular models being discussed in the literature for the disruption of star clusters over the first approximately a few hundred million years of their lives, one where lower mass clusters are disrupted earlier than their higher mass counterparts (mass-dependent disruption---e.g., Bastian et~al.\ 2012) and one where clusters disrupt 
at approximately the same rate, regardless of their mass (mass-independent disruption---e.g., Fall \& Chandar 2012). The models make different predictions for the shape of the age distribution in different intervals of mass and for the mass function in different intervals of age. Basic predictions from each model are summarized below, and then compared with the observed mass-age distributions of star clusters in our two M83 fields.

In gradual, mass-dependent disruption (MDD) models (e.g., Boutloukos \& Lamers 2003; Fall et~al.\ 2010), clusters lose mass at approximately the same rate, leading to the earlier disruption of lower mass clusters when compared with their higher mass counterparts.  
This model predicts `breaks' or curvature in the mass (and age) distributions for a population
of clusters. In this model the disruption time $\tau_d$ has been characterized as 
$\tau_d(M) = \tau_* (M/M_*)^k$, where the exponent $k$ and characteristic disruption
timescale $\tau_*$ are adjustable parameters, and $M_*=10^4~M_{\odot}$ is a fiducial mass scale (Lamers et~al.\ 2005; Fall et~al.\ 2009). We first assume that the initial shape of the mass function is a power-law, and that clusters form at a constant rate. Mass-dependent disruption models predict that the cluster age distribution (for mass-limited samples, as presented here), will be flat at young ages, but then fall off exponentially at an age that reflects the characteristic disruption time. This behavior should occur at all mass ranges, but the break point will occur at younger ages for lower mass clusters. The mass function for the youngest clusters will have a power-law shape, which will flatten towards lower masses at older ages if mass-dependent disruption affects the clusters. The reader is referred to Figures~10 and 11 in Fall et~al.\ (2009) for graphical examples of these predictions. The predicted behavior of the mass function is similar if the initial shape is a Schechter function rather than a power-law; an example is shown in Figure~8. For both assumed initial distributions, \emph{a critical prediction of mass-dependent disruption models is the flattening of the mass function at older ages and lower masses.}

In the gradual, mass-independent disruption model the two distributions are independent of one another, and can be written as:  $g(M,\tau) \propto M^{\beta} \tau^{\gamma}$.
This model predicts a power-law decline in the number of clusters at each mass with age, at a fractional rate that is independent of their masses. The age distribution declines as a power law in each interval of mass. Mass-independent disruption models predict that there should be no change in the shape of the cluster mass function, i.e., no flattening occurs at the low mass end. Again, the reader is referred to Figure 12 in Fall et~al.\ (2009).

\subsection{Comparison between Predictions and Observations}

We first compare our observed distributions with predictions from 
the mass-\emph{dependent} disruption model. The age distributions (Figure~5) in the inner field,  which are only plotted for $\tau \gea 10$~Myr, are inconsistent with this model (i.e. no curvature is observed), and can be reasonably well represented by a single power-law with $\gamma\approx -0.7$ to $-0.9$, although the log~$(\tau/\mbox{yr}) =7.0$--7.5 bin does appear low in cases where a gap due to the age-dating artifact (mentioned in 
Section~3.1) is present. The outer field age distributions are relatively similar to those of the inner field, with any difference such that the outer field is shallower occurring at the 
$\approx 2$--$3\sigma$ level (see Section~3.2).

More importantly, the shape of the mass functions in the inner and outer fields do not 
flatten over time. Figure~7 shows this explicitly, i.e. the red squares, showing 100--400~Myr-old clusters, do not deviate below the younger age populations (green triangles and blue circles) at masses around and below $10^4M_{\odot}$. This is true regardless of whether or not there is a downturn at the high mass end. Therefore, any dependence of the disruption rate of the clusters on their mass, if it exists, is too weak to be observed, even with the high quality observations taken with the \emph{HST}. \emph{We conclude that the M83 clusters studied here, in both the inner and outer fields, do not show evidence for mass-dependent disruption over the observed $M - \tau$ domain.}

Next, we compare our observed age and mass distributions with predictions from the 
mass-\emph{independent} disruption model. The observed mass and age distributions in both fields are \emph{consistent} with predictions from this model,  i.e., the age distributions are well described by a single power law that is approximately independent of cluster mass, and the mass function can be described by a power-law that is approximately independent of the age of the clusters. The main difference in the results between the two fields is in the exponent $\gamma$, \textbf{with the inner field having $\gamma\approx -0.7$ to $-0.9$ } and the outer field having $\gamma\approx -0.5$. Adopting realistic uncertainties and bracketing a range of possible differences in the relative star formation histories between the two fields (from no difference to the maximum suggested by 
Figure~11 in Bastian et~al.), we find that the exponents only differ at the $\approx$2--$3\sigma$ level, as discussed in Section~3.2. We also note that a $\gamma$ value of $-0.5$ still indicates strong cluster disruption, with $\approx$70\% of clusters disrupting every decade in age  (i.e., (1--$10^{-0.5}) \times 100\% = 68\%$), not very different from the approximately 80--90\% disruption suggested by Chandar et~al.\ (2012), or found in this paper for the inner field (i.e., (1--$10^{-0.8}) \times 100\% = 84\%$).

\subsection{Agreement and Disagreement with Previous Interpretation}

Our results are quite similar, in most regards, to those found by Bastian et~al.\ (2012), and are now supported by the addition of two cluster catalogs in each field selected using different methods. Both groups find similar looking color-color diagrams (Figure~3), similar mass-age diagrams (Figure~4), and similar age and mass distributions (Figures 5--7), except for some differences at $\tau \lea 10^7$~yr, as expected due to differences
in the selection criteria (see Section~2.2 here and Bastian et~al.\ 2012). Both groups agree that mass-independent disruption provides a reasonable description of the data, and that mass-dependent disruption models that assume an initial power-law mass function do 
\emph{not} fit the data well.

The biggest area of disagreement is the contention by Bastian et~al.\ (2012) that mass-dependent disruption models that assume an initial Schechter mass function 
\emph{can} also fit the data. When they perform a two-dimensional fit to clusters in the $M-\tau$ plane, they derive the specific values for $\tau_*$ and $M_C$ of 160~Myr and
$1.5\times 10^5~M_{\odot}$ in the inner field, and 600~Myr and $5\times10^4~M_{\odot}$ in the outer field. In Figure~\ref{fig:dndlogm_datamod} we compare the observed 
mass functions of 100--400~Myr clusters in both fields with predictions from the Bastian 
et~al.\ (2012) mass-dependent disruption model. The panels on the left show the predicted evolution assuming their best fit model parameters in each field. The blue dashed lines show the initial Schechter mass function, the green dotted and red solid lines show the predicted evolution for 10--100~Myr and 100--400~Myr cluster populations, respectively. The shorter disruption time $\tau_*$ in the top-left panel leads to faster evolution and more flattening at the low mass end of the cluster mass function when compared with the bottom-left panel. The panels on the right in Figure~\ref{fig:dndlogm_datamod} compare
the observed mass function for 100--400~Myr clusters with the model predictions. Here, we have allowed the flexibility of renormalizing the predicted 100--400~Myr distribution to best match the shape of the observed distribution, by matching the predictions and observations at the high mass end. However, even with this added degree of flexibility, the specific mass-dependent disruption model and parameters suggested by Bastian et~al.\ (2012) do not provide a good match to the observations, i.e., the red curves are clearly flatter at the low mass end than the observed ones. In fact, the initial Schechter function (blue curve), which represents no mass-dependent disruption, provides a much better fit to the observations. Therefore, mass-dependent disruption cannot have much affect on the 
observed $M-\tau$ ranges of the cluster population in these two fields of M83. Any mass-dependent disruption, if it exists,  must occur below the selection limits of these catalogs.

\section{Summary and Conclusions}

In this paper we determined the mass and age distributions of star clusters detected in two fields observed with the \emph{HST}/WFC3 in the nearby spiral galaxy M83, and compared them with predictions from two different models of cluster disruption. We used three distinct catalogs in each field for this purpose, including one from the previously published work by Bastian et~al.\ (2012), where the clusters were selected using different methods and criteria. In each case,  integrated \textit{UBVI}~H$\alpha$ photometric measurements were compared with predictions from population synthesis models in order to estimate the age ($\tau$) and mass ($M$) for each cluster.

We found that the age and mass distributions, $dN/d\tau$ and $dN/dM$, of the clusters in each field did not differ significantly amongst the catalogs, particularly for $\tau \gea 10^7$~yr. These distributions are reasonably described by single power laws, $dN/d\tau \propto \tau^{\gamma}$ and $dN/dM \propto M^{\beta}$.  We found $\gamma \approx -0.84\pm0.12$ and $\beta \approx -1.98\pm0.14$, for the inner field, and $\gamma \approx -0.48\pm0.12$ and $\beta \approx -2.34\pm0.26$ for the outer field. The relative difference between the star formation histories in the two fields are uncertain, but result in a range $\Delta \gamma = 0.18 - 0.36 \pm 0.12$, i.e., if the $\gamma$ values between the two fields differ,
it is at the 2--3$\sigma$ level. We concluded that the shapes of the mass and age distributions of the clusters in the two fields are similar, to first order, as predicted by the ``quasi-universal'' model, although it is possible that other dependencies may play a weak role.

The shapes of the cluster age distributions were roughly independent of mass, and the shapes of the cluster mass functions were approximately independent of age, at least over
the studied $M-\tau$ range. In addition, none of the distributions showed any obvious curvature at lower masses or older ages. Our results are consistent with the clusters being disrupted, starting soon after they form,  at a rate that is approximately independent of their mass. Our results do not show evidence of mass-dependent disruption, where lower mass clusters are disrupted earlier than their higher mass counterparts. In a future study, we will include observations of clusters in five additional pointings within M83, observed with the WFC3 camera on \emph{HST}, to investigate the disruption histories of the clusters in more detail.

\acknowledgments

We thank Zolt Levay for making the color images used in Figures~1 and 2, and Mike Fall, Nate Bastian, and the referee for helpful comments on an earlier version of the manuscript.
This paper is based on observations taken with the NASA/ESA \emph{Hubble Space Telescope} obtained at the Space Telescope Scinece Institute, which is operated by AURA, Inc., under NASA contract NAS5-26555. It uses Early Release Science observations made by the WFC3 Science Oversight Committee, and was supported in part by STScI grant GO-11359. We are grateful to the Director of STScI for awarding Director's Discretionary time for this program. This research has made use of the NASA/IPAC Extragalactic
Database (NED), which is operated by the Jet Propulsion Laboratory, California Institute of Technology, under contract with NASA.

{\it Facilities:} \facility{HST}.

\begin{figure}
\hspace{1in} \includegraphics[height=7in]{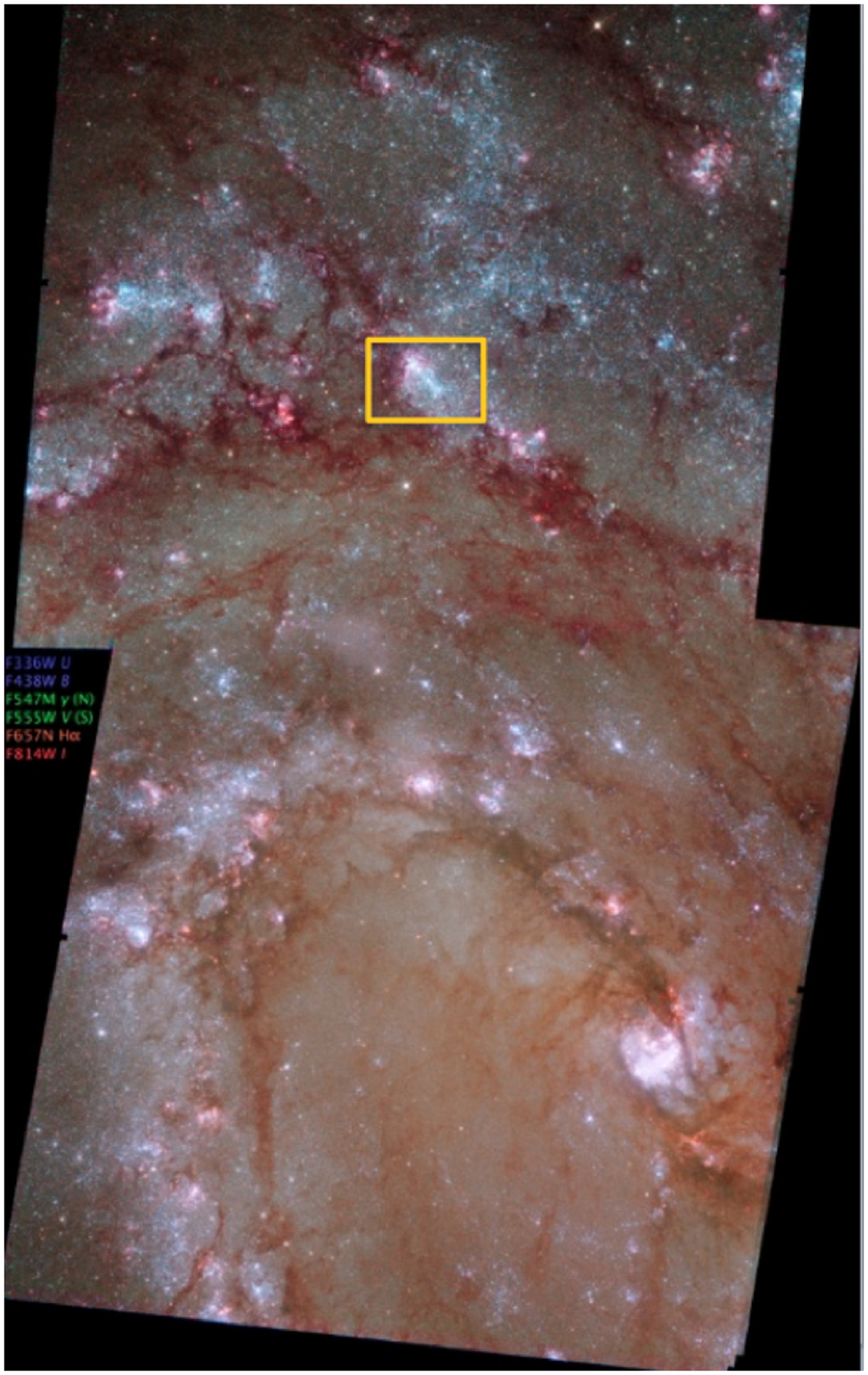}
\caption{Color images of two fields in M83 produced using the \emph{HST}/WFC3 observations described in this work.  $U$ plus $B$-band images are shown in blue,
the $V$ band in green, and a combination of the $I$ and H$\alpha$ filters in red.  The rectangular region is used in Figure~2 to compare different cluster catalogs.
}
\label{fig:prettypics}
\end{figure}

\begin{figure}
\epsscale{0.8}
\plotone{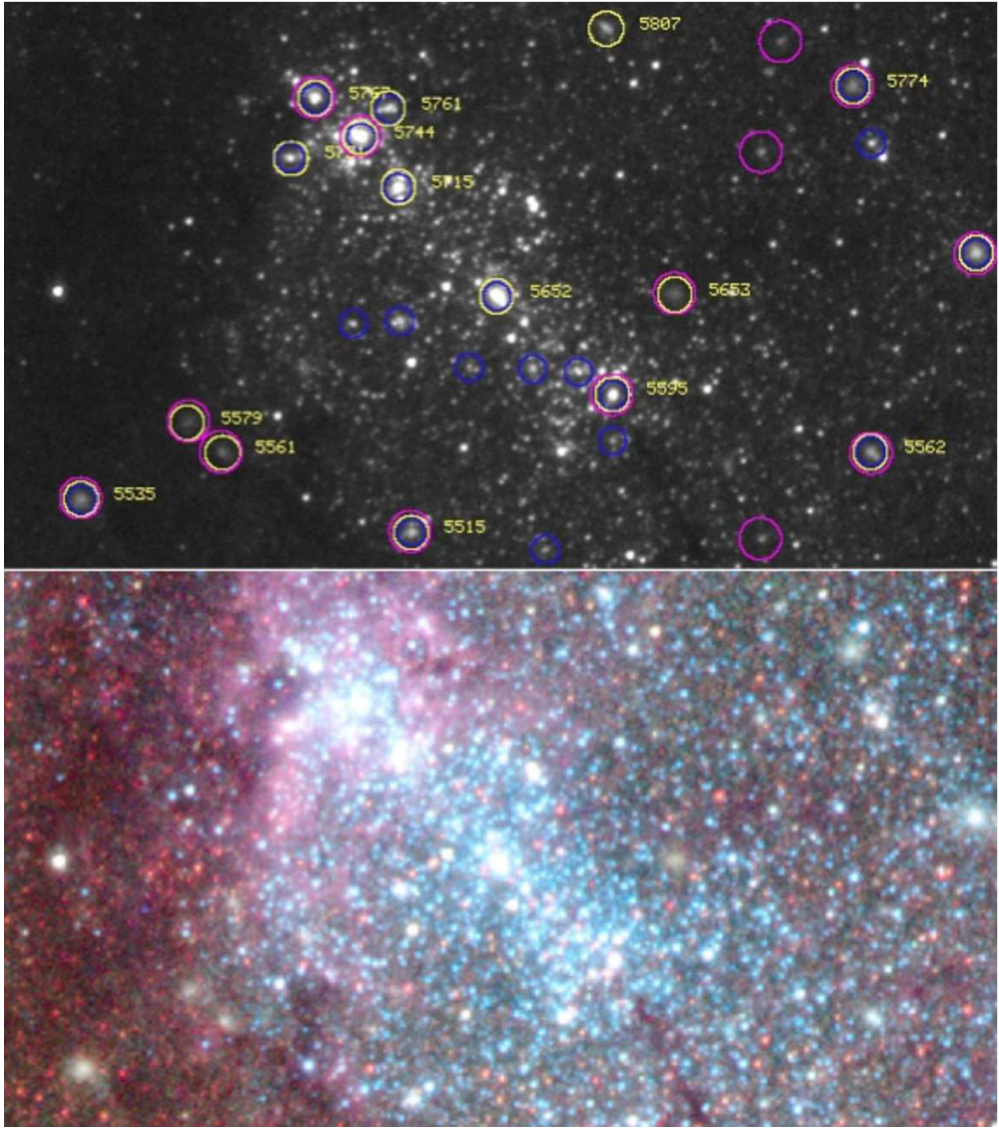}
\caption{The upper panel shows a comparison of clusters selected in the portion of the outer field highlighted in Figure~\ref{fig:prettypics} in the automated (blue circles) and manual (yellow circles) catalogs presented in this work, and in the cluster catalog presented by Bastian et~al.\ 2011 (magenta circles). In this particular region, the catalogs agree at the 
40--70\% level.  Most of the differences between the catalogs occur for very young $\tau \lea 10^7$~yr clusters (e.g., objects 5652 and 5715 are not included in the Bastian et~al.\ catalog), which tend to reside in the most crowded regions. The lower panel shows a color image of the same region.
}
\label{fig:catcomp1}
\end{figure}

\begin{figure}
\begin{center}
\includegraphics[width = 3in, angle= -90]{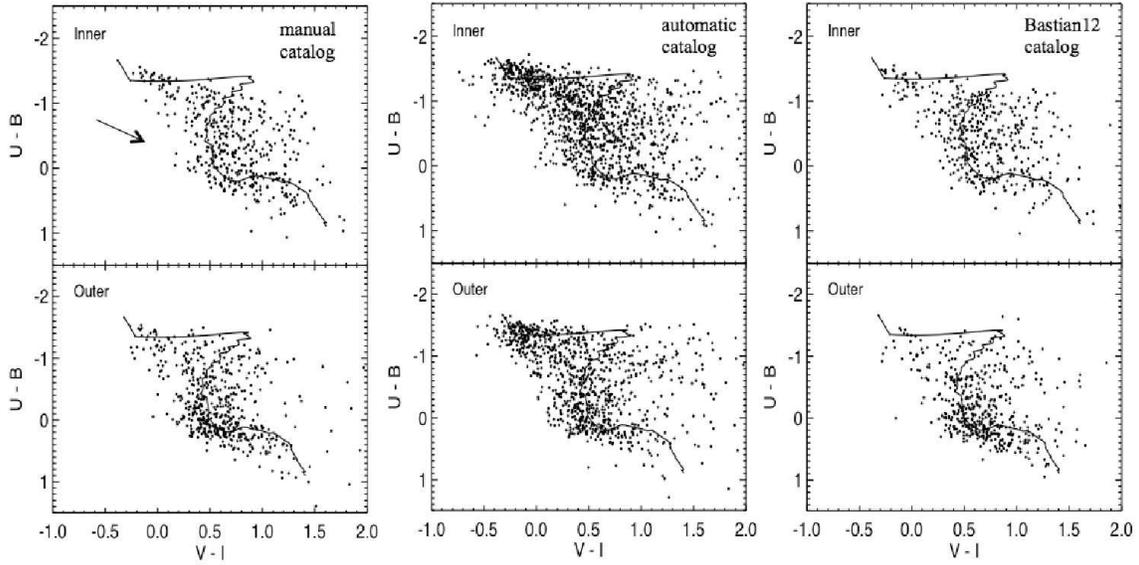}  
\end{center}
\caption{($U-B$) vs.\ $(V-I)$ two-color diagram for cluster candidates in the manual (left), 
automatic (middle), and Bastian et~al.\ (2012, right) catalogs, as discussed in the text.
Catalogs for the inner field are shown in the upper panels, and the outer field in the lower panels. The curves show predictions in the appropriate WFC3 filters from the single stellar
population models of Charlot \& Bruzual (2009, private communication) for twice solar (solid line) metallicity. Ages range from  $10^6$~yr at the upper left end of the curve to 
$10^{10}$~yr at the bottom right end. An  $A_V = 1.0$ reddening vector is shown in the upper left panel. }
\label{fig:2color}
\end{figure}

\begin{figure}
\begin{center}
\includegraphics[width = 3in, angle= -90]{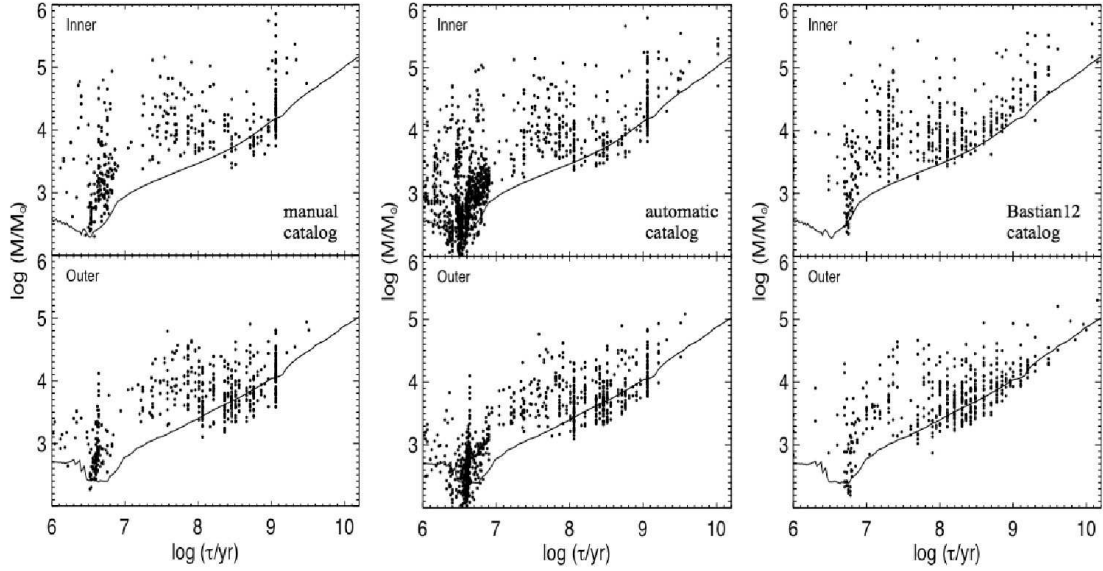}  
\end{center}
\caption{Log~$M$ vs.\ log~$\tau$ derived for clusters in the indicated catalogs. The solid line in each panel shows a magnitude limit of $M_V=-6.0$.
}
\label{fig:Mt}
\end{figure}

\begin{figure}
\begin{center}
\includegraphics[width = 3in, angle= -90]{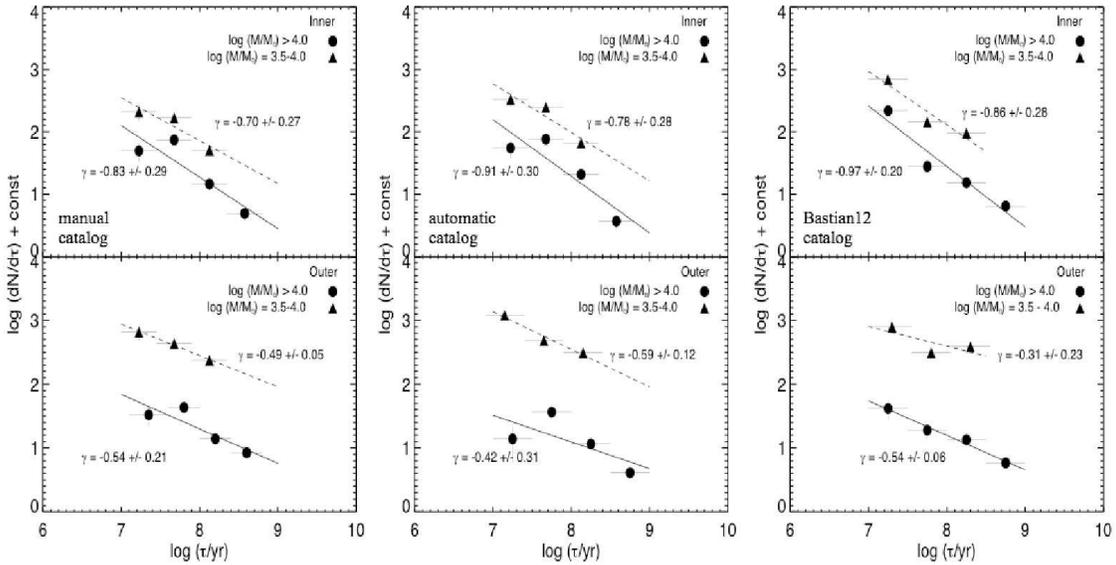}  
\end{center}
 \caption{Age distribution of star clusters in M83 in different mass intervals using the indicated cluster catalogs. The normalizations of the age distributions are arbitrary. The lines show power laws, $dN/d\tau \propto \tau^{\gamma}$, with the best-fit exponents listed in Table~1. 
 }
 \label{fig:dndt}
 \end{figure}

 \begin{figure}
\begin{center}
\includegraphics[width = 3in, angle= -90]{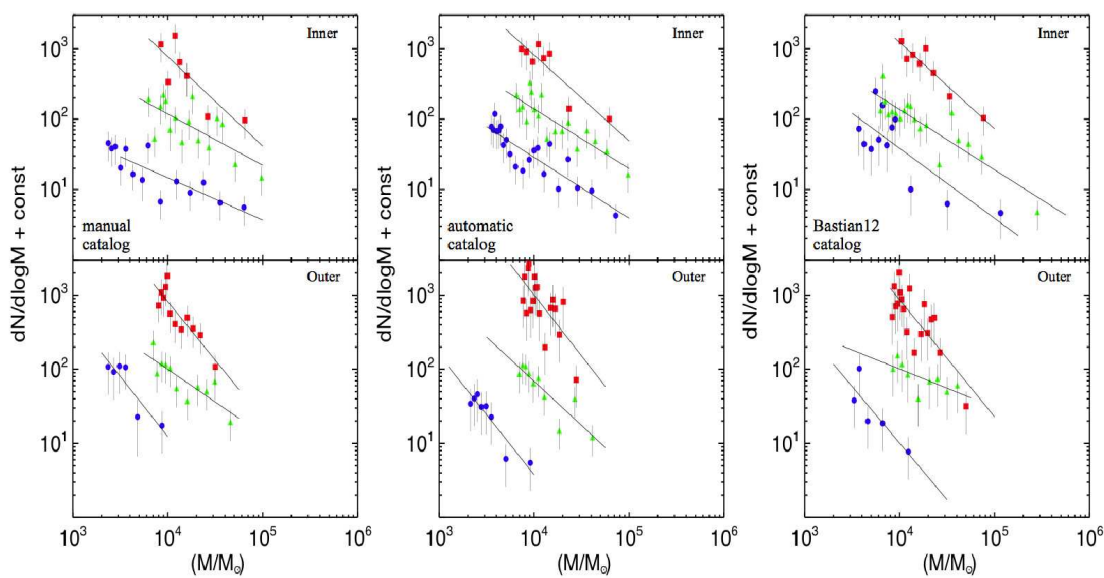}
\end{center}
 \caption{Mass functions of star clusters in M83 in different age intervals using the indicated cluster catalogs. The data are restricted to stay above the approximate completeness
limit, shown as the solid line in Figure~\ref{fig:Mt}. The lines show power laws, $dN/dM \propto M^{\beta}$, with the best fit exponents listed in Table~2. The color code is: 
log~$\tau=6.0$--7.0 (blue), log~$\tau=7.0$--8.0 (green), and log~$\tau=8.0$--8.6 (red).
 }
 \label{fig:dndlogm}
 \end{figure}

 \begin{figure}
\begin{center}
\includegraphics[width = 3in, angle= -90]{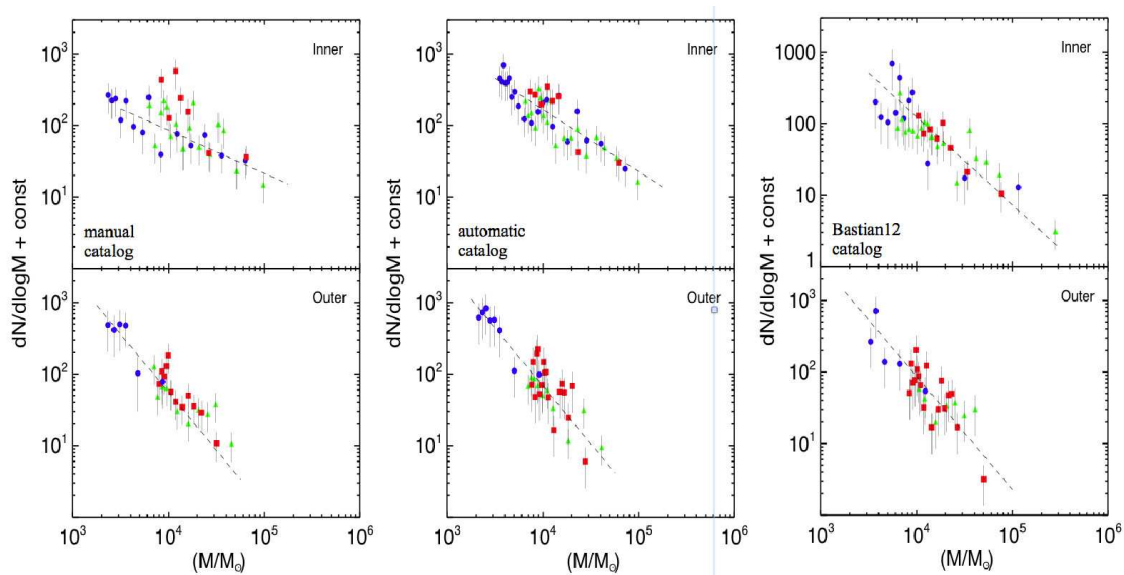}  
\end{center}
 \caption{Mass functions of star clusters in M83 in different  age intervals using the indicated cluster catalogs.  The color coding for the ages are the same as in Figure~6.  The dashed line shows a power law that approximately matches the shape of the cluster mass function in each catalog. Each distribution has been offset by an arbitrary normalization so that they lie on top of one another, in order to facilitate comparison of the shapes of the mass functions at different ages.  There is no flattening for the mass functions at the oldest ages
(red data points) in any of these panels.
 }
 \label{fig:dndlogmnorm}
 \end{figure}

 \begin{figure}
 \epsscale{0.65}
\plotone{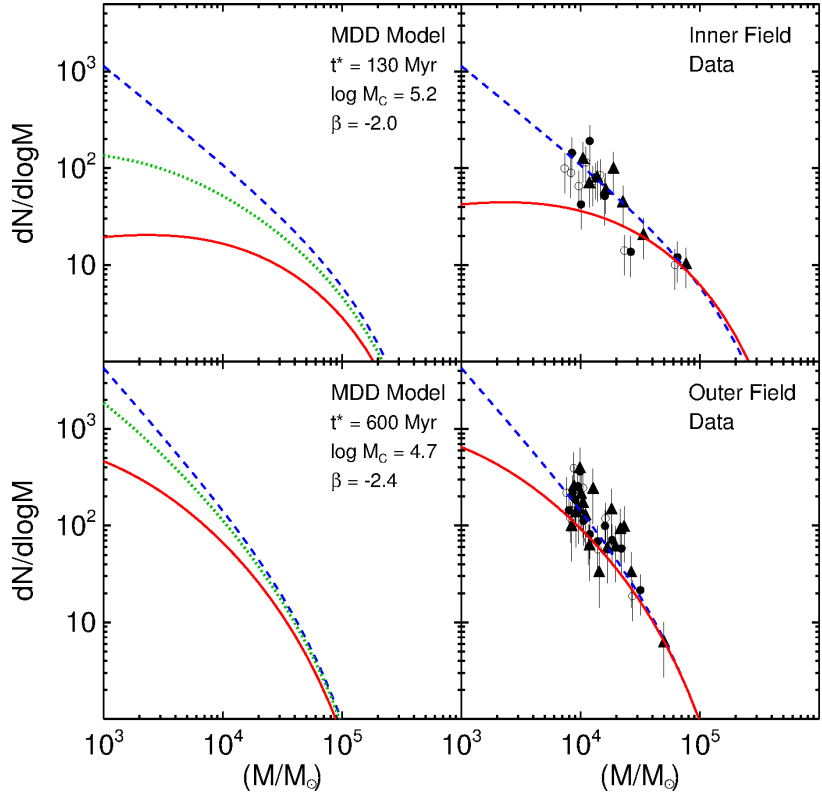}
 \caption{The panels on the left show the predicted evolution of the cluster  mass function based on the mass-dependent disruption (MDD) model  used by Bastian et~al.\ (2012).
The model assumes an initial Schechter function, $dN/dM \propto M^{\beta} \mbox{exp}^{M/M_C}$, that is evolved using the formalism of Lamers et~al.\ (2005), where the disruption timescale is given by $\tau_{dis}(M) = \tau_* (M/M_*)^{k}$. Model curves of the predicted mass function for three different population ages are shown in each panel:
the initial distribution (dashed blue line, appropriate for 1--10~Myr clusters), 10--100~Myr (green dotted line) and 100--400~Myr (solid red line). The model parameters, $\tau_*$, 
$M_C$, and $\beta$, are those determined by Bastian et~al.\ (2012) as the  best fit to the clusters in the inner and outer field of M83,  assuming $k=0.65$. The panels on the right compare the observed mass functions for  100--400~Myr clusters in M83 from the manual (filled circles), automatic (open circles), and hybrid (filled triangles) catalogs, with two of the predicted curves shown on the left. The model curves for the 100--400~Myr population are renormalized here, in order to best match the shape at the upper end of the observed distribution. The key result is that the predicted mass-dependent evolution (red curves)
with the parameters determined by Bastian et~al.\ (2012) are a poor match to the observations. The blue curve, representing \emph{no} mass-dependent disruption, provides a
better description of the data.
}
 \label{fig:dndlogm_datamod}
 \end{figure}

\clearpage
 
 \begin{deluxetable}{lcc}
\tablecolumns{3}
\tablecaption{Exponents of Age Distributions\label{gammavalues}}
\tablewidth{0pt}
\tablehead{
\colhead{Cluster} & \colhead{Mass Range}  & \colhead{$\gamma$\tablenotemark{1}} \\
\colhead{Catalog} & \colhead{log($M/M_{\odot}$)}  & \colhead{}
}
\startdata
Inner Field, manual  & $>$\,4.0 & $-0.83 \pm 0.29$\\
Inner Field, automatic &  $>$\,4.0 & $-0.91\pm0.30$\\
Inner Field, Bastian & $>$\,4.0 & $-0.97 \pm 0.20$\\ \hline
Inner Field, manual  & 3.5--4.0 & $-0.70 \pm 0.27$\\
Inner Field, automatic & 3.5--4.0 & $-0.78\pm0.28$\\
Inner Field, Bastian & 3.5--4.0 & $-0.86 \pm 0.28$\\ \hline
Outer Field, manual  & $>$\,4.0 & $-0.54 \pm 0.21$\\
Outer Field, automatic &  $>$\,4.0 & $-0.42 \pm 0.31$\\
Outer Field, Bastian & $>$\,4.0 & $-0.54 \pm 0.06$\\ \hline
Outer Field, manual  & 3.5--4.0 & $-0.49 \pm 0.05$\\
Outer Field, automatic & 3.5--4.0 & $-0.59 \pm 0.12$\\
Outer Field, Bastian & 3.5--4.0 & $-0.31 \pm 0.23$\\
\enddata
\tablenotetext{1}{\null Least-squares fits to log$(dN/d\tau) = \gamma\log\tau + 
\mathrm{ const}$}
\end{deluxetable}

\clearpage
\begin{deluxetable}{lcc}
\tablecolumns{3}
\tablecaption{Exponents of Mass Functions\label{betavalues}}
\tablewidth{0pt}
\tablehead{
\colhead{Cluster}  & \colhead{Age Range} & \colhead{$\beta$\tablenotemark{1}} \\
\colhead{Catalog} & \colhead{log($\tau$/yr)} & \colhead{}
}
\startdata
Inner Field, manual  & 6--7\phm{.6} & $-1.70 \pm 0.12$\\
Inner Field, automatic &  6--7\phm{.6} & $-1.87\pm0.11$\\
Inner Field, Bastian & 6--7\phm{.6} & $-2.00 \pm 0.26$\\ \hline
Inner Field, manual  & 7--8\phm{.6} & $-1.84 \pm 0.19$\\
Inner Field, automatic & 7--8\phm{.6} & $-1.89\pm0.12$\\
Inner Field, Bastian & 7--8\phm{.6} & $-1.86 \pm 0.11$\\ \hline
Inner Field, manual  & 8--8.6 & $-2.24 \pm 0.23$\\
Inner Field, automatic &  8--8.6 & $-2.21\pm0.23$\\ 
Inner Field, Bastian & 8--8.6 & $-2.22 \pm 0.19$\\ \hline
Outer Field, manual  & 6--7\phm{.6} & $-2.59 \pm 0.39$\\
Outer Field, automatic &  6--7\phm{.6} & $-2.61\pm0.30$\\
Outer Field, Bastian & 6--7\phm{.6} & $-2.49 \pm 0.15$\\ \hline
Outer Field, manual  & 7--8\phm{.6} & $-1.88 \pm 0.20$\\
Outer Field, automatic & 7--8\phm{.6} & $-2.19\pm0.23$\\
Outer Field, Bastian & 7--8\phm{.6} & $-1.54 \pm 0.18$\\ \hline
Outer Field, manual  & 8--8.6 & $-2.58 \pm 0.27$\\
Outer Field, automatic &  8--8.6 & $-2.64\pm0.44$\\ 
Outer Field, Bastian & 8--8.6 & $-2.58 \pm 0.32$\\
\enddata
\tablenotetext{1}{\null Least-squares fits to log$(dN/dM) = \beta\log{M} + 
\mathrm{const}$}
\end{deluxetable}


\begin{thebibliography}{}

\bibitem{ref1}
Bastian, N., Adamo, A., Gieles, M., Silva-Villa, E., Lamers, H.~J.~G.~L.~M., Larsen, 
S.~S., Smith, L. J., Konstantopoulos, I. S., Zackrisson, E. 2012, MNRAS, 419, 2606

\bibitem{ref1}
Baumgardt, H., \& Kroupa, P. 2007, MNRAS, 380, 1589

\bibitem{ref1}
Boutloukos, S.~G., \& Lamers, H.~J.~G.~L.~M. 2003, MNRAS, 338, 717

\bibitem{ref1}
Bruzual, G., \& Charlot, S. 2003, MNRAS, 344, 1000

\bibitem{ref1}
Chabrier, G. 2003, PASP, 115, 763

\bibitem{ref1}
Chandar, R., Fall, S.~M., \& Whitmore, B.~C. 2010, ApJ, 719, 996

\bibitem{ref1}
Fall, S.~M., Chandar, R., \& Whitmore, B.~C. 2005, ApJ, 631, L133

\bibitem{ref1}
Fall, S.~M., Chandar, R., \& Whitmore, B.~C., 2009, ApJ, 704, 453

\bibitem{ref1}
Fall, S.~M., \& Chandar, R. 2012, ApJ, 752, 96

\bibitem{ref1}
Fall, S.~M., \& Zhang, Q. 2001, ApJ, 561, 751

\bibitem{ref1}
Fitzpatrick, E.~L. 1999, PASP, 111, 63

\bibitem{ref1}
Fouesneau, M., Lancon, A., Chandar, R., \& Whitmore, B.~C. 2012, ApJ, 750, 60

\bibitem{ref1}
Koekemoer, A.~M., Fruchter, A.~S., Hook, R.~N., \& Hack, W. 2002, in The 2002 HST Calibration Workshop, MultiDrizzle: An Integrated Pyraf Script for Registering, Cleaning, and Combining Images, ed.\ S.~Arribas, A.~Koekemoer, \& B.~Whitmore (STScI: Baltimore), 337

\bibitem{ref1}
Lamers, H.~J.~G.~L.~M., Gieles, M., Portegies Zwart, S.~F. 2005 A\&A, 429, L173

\bibitem{ref1}
Miller, B.~W., Whitmore, B.~C., Schweizer, F., \& Fall, S.~M. 1997, AJ, 114, 2381

\bibitem{ref1}
Salpeter, E. 1955, ApJ, 121, 161

\bibitem{ref1}
Silva-Villa, E., Adamo, A. \& Bastian, N. 2013, MNRAS, 436, 69

\bibitem{ref1}
Silva-Villa, E., \& Larsen, S.~S. 2011 A \& A 529, 25

\bibitem{ref1}
Thim, F., Tammann, G.~A., Saha, A., Dolphin, A., Sandage, A., Tolstoy, E., \& Labhardt, L. 2003, ApJ, 590, 256

\bibitem{ref1}
Whitmore, B.~C., Chandar, R., \& Fall, S.~M. 2007, AJ, 133, 1067

\bibitem{ref1}
Whitmore, B.~C., et al. 2013, submitted to AJ

\end{thebibliography}
\end{document}